 \documentclass{aastex}

\begin{document}

\title{STAR FORMATION HISTORY
IN THE SOLAR VICINITY}

\author{G. Bertelli}
\affil{Consiglio Nazionale delle Ricerche, CNR-GNA, Roma, Italy }
\affil{and Dipartimento di Astronomia, Universit\'{a} di Padova, Vicolo 
dell'Osservatorio, 5, I-35122 Padova, Italy}
\email{bertelli@pd.astro.it}

\author{E. Nasi}
\affil{Osservatorio Astronomico di Padova, Vicolo dell'Osservatorio,
5, I-35122 Padova, Italy}
\email{nasi@pd.astro.it}

\begin{abstract}

The star formation history in the solar neighbourhood is inferred comparing a 
sample of field stars from the Hipparcos Catalog   
with synthetic CMDs.  We considered separately the main sequence and
the  red giant region of the HR diagram. The criteria for our 
best solutions are based on the  $\chi^{2}$  minimization  
of star distributions in  selected zones of the HR diagram.
Our analysis suggests that: a) the solutions are compatible with a Salpeter IMF
and with {\sl a star formation rate increasing, in a broad sense, from
the beginning  to   the present time}; 
b) the deduced volume mass densities and the corresponding absolute scale of
the SFR solutions are strongly influenced by the initial mass function slope
of low mass stars (below $0.5 M_{\odot}$); 
c) the stellar evolutionary models are not 
completely adequate: in fact {\sl the theoretical ratio 
between the   He-burning  and  MS star numbers 
is always a factor  1.5  greater than the observational value}. 
This fact could indicate 
the need of a more efficient overshoot in the evolutionary models, or
a different mixing theory. 

\end{abstract}


\section {Introduction}

Determining the past history of star formation from the Color-Magnitude
Diagram (CMD) of composite stellar populations in galaxies of different 
morphological type is one of the main targets of modern astrophysics.
For nearby galaxies, in which individual stars are resolved and CMDs
are derived, the problem is easier to be tackled
as all stars are placed nearly at the same distance. However
in our own galaxy the problem is by far more complicated because
there are large fractional differences in the distances of the galactic stars
 and only CMDs containing stars of non homogeneous age, chemical
composition and distance are available. The evaluation of the underlying star 
formation history (SFH) requires a few assumptions besides
postulating a suitable star formation rate (SFR) when  comparing
theoretical predictions with observational data. For the sake of 
illustration,
let us consider the case of chemical models designed to interpret the chemical
abundances and their spatial gradients in the Galactic Disk. In these models 
the SFR is assumed to be first increasing and then decreasing up to the 
present time; see for instance
model A of Chiappini et al. (1997). But other independent  
studies suggest that the local SFR has been 
quite irregular, with periods of enhancement and quiescence showing
significant fluctuations (see Rocha-Pinto et al. 1996, 1997, 2000 and 
references therein). Using the chromospheric activity-age distribution of 
stars in the solar neighbourhood, 
they found a burst of  star formation  8 Gyr ago, and a lull of activity
some  2-3 Gyr ago for the Milky Way.   However, 
the application of their method to the data of Edvardsson et al. (1993)
produces a SFR smoothly increasing up to the present.

Recently  Hernandez et al. (2000) studied the star formation history
of the Hipparcos solar neighbourhood with a maximum likelihood statistical
approach combined with variational calculus. They derived an oscillatory
component of period $\sim 0.5$ Gyr, superimposed on a small level of
constant star formation activity and the involved time interval covers the 
last 3 Gyr. The restriction to
this small age range is compelled by the selection criteria (involving 
limiting magnitude and distance error of the Hipparcos sample) required 
by their method. 

The advent of the Hipparcos mission set a landmark, as 
for the
first time it was possible to derive the CMD of field stars in the solar
vicinity based on accurate distances for
each   individual object (Perryman et al. 1995). 
The solar neighbourhood sample is very useful for  studying  our Galaxy's
past history, as it is
characterized by a large range of ages, distances and metallicities.
In addition to the past SF history (see Bertelli et al. 1997, 1999),
the Hipparcos data of field stars have been the subject of several studies
aimed at testing stellar evolution theory (see Schr\"oder 1998, 
Girardi 1999). 

It is the purpose of this paper to present the results of our investigation
for the star
formation history of the solar neighborhood on a selected sample of 
Hipparcos data.

\section {The Local Solar Neighborhood Sample of  Field Stars  }

In figure 1a we present the HR diagram for our sample of field stars
selected from  the Hipparcos Catalog
with the criteria described in section 2.4 below. 
In figure 1b 
we define several regions where we compare the theoretical and 
observational stellar distributions    
to test different hypotheses for the SFR under the criteria for chi-square
minimization.   
 The use of characteristic HRD regions has a relevant
advantage since star counts are insensitive to small errors in
luminosity,  colour and  small differences in
chemical abundances,
provided the sample be adequately populated. 

\subsection {Completeness}

The evaluation of the  completeness for our stellar sample is fundamental 
for any statistical analysis.
According to Turon et al. (1992) the Hipparcos Catalogue is essentially 
complete 
up to $ V \sim 9$ mag and, as reported by Perrymann et al. (1995), the  
limiting magnitude is determined by the sensitivity threshold of its starmapper 
detectors 
and depends on the galactic latitude and the specral type ($V=7.9+1.1sin|b|$
for types earlier than G5  and $V=7.3+1.1sin|b|$ for later spectral types).

Schr$\ddot{o}$der (1999) discussed the completeness of the stellar sample based
on Hipparcos data in his paper on the solar neighbourhood HR diagram as a test
of the evolutionary timescales.
For an HRD of stars within $d = 100$ pc of the Sun, the sample can be 
considered complete down to $M_V = 4.0$;  within $d = 50 $ pc 
completeness is expected to about $M_V = 5.5$

As the completeness limit depends on the latitude, we
restricted the subsequent counts to HRD regions with $M_V$ brighter than 4.5
to have a reliable completeness down to the low luminosity end of the
considered 50 pc sample.

\subsection {Statistical bias}

If stars are selected according to the parallax value ( $\pi >
 \pi_{threshold}$ ) and/or if weights are given according to the ratio between 
its error and the measured parallax 
($\sigma_{\pi}/\pi$), parallaxes are systematically
overestimated and  the stars in a volume limited sample are on average 
brighter than they appear to be. 
The corresponding corrections are called Lutz-Kelker corrections (Lutz 
\& Kelker 1973). They must be applied to  absolute
magnitudes  and may be evaluated by means of 
equation 31 in Hanson (1979), where the parallax distribution is
characterized as a power law,
$P(\pi) \propto \pi^{-n}$ (in case of a uniform space distribution n=4). 
If the stellar sample under study is wholly or partially magnitude-limited,
then one expects $n<4$, which means lower predicted corrections.  With fewer 
stars at smaller parallax, the probability of overestimating an individual
parallax measurement is correspondingly reduced (Reid 1997).
The resultant Lutz-Kelker corrections strongly depend on the parallax
uncertainty, so that the accuracy of Hipparcos parallaxes (we select stars 
with $ \pi > 20$ mas and $\sigma_{\pi} / \pi < 0.1 $ ) limits the LK bias to 
magnitude corrections of the order of 0.1 magnitudes at maximum for the case 
$n=4$ of uniform space distribution. The safest procedure is to use only
stars with good parallaxes ($\leq 10 \%$ accuracy), since corrections are 
negligible for the Hipparcos sample considered in our analysis.  
If stars have a range of luminosities, a magnitude limited sample would be 
biased toward luminous objects (Malmquist bias) as the average absolute
magnitude is systematically overestimated. The corrections for Malmquist
bias have the  opposite sign with respect to Lutz-Kelker corrections (they may
even compensate each other).
 
\subsection {Binaries}

The high incidence of binaries among field stars (possibly as high as $60\%$ )
is a well established fact (Pont et al. 1998). When both components have 
similar luminosities, an unresolved binarity causes an apparent increase of
up to 0.75 magnitudes. Owing to the sharp dependence of the Hipparcos parallax
uncertainty on magnitude (Pont et al. 1998), a binary star of a given 
colour has a higher likelihood (than a single star) of being included in a 
sample selected by $\sigma_{\pi}/\pi$ limits .

It is necessary to take into account the distribution of binary system mass 
ratios, if we want to introduce binaries in our HRD simulations  of the
Hipparcos sample. There are several papers dealing with these problems.
As far as mass ratios are involved, Trimble (1990), Mermilliod et al. (1992)
and Mazeh et al. (1992) for example, considered different samples of 
spectroscopic binaries and found different forms for the distribution function
of the mass ratios, probably due more to the selection of binary systems than 
to the method of analysis. Defining $q = M_2 / M_1 $, 
the distribution N(q) is best fitted by a power law near to $q^{-1}$ over the 
range $ q = 0.1 - 1.0$, according to Trimble (1990). A similar declining power
law has been found for Hyades binaries by Patience et al. (1998), pointing 
out that there is no mass dependence of the mass ratio distribution. 
   From spectroscopic binaries in the Pleiades Mermilliod et al. (1992)
gave a mass ratio distribution reasonably flat for q between 0.4 and 1.0 and
no information for the behaviour for $q < 0.4$.

There is also a variation of the proportion of binaries with the mass of 
the primary, as summarized by Kroupa (1995), going from 0.42 for M dwarfs
and 0.45 for K dwarfs to 0.53 for G dwarfs. Duquennoy and Mayor (1991), in their
paper on multiplicity among solar-type stars in the solar neighbourhood,
found that only about one third of the G-dwarf primaries may be real single
stars. 
The multiplicity of the massive stars in the Orion Nebula cluster is quite
high ( of the order of 1.5 companions per primary star on average, after 
correction for unresolved systems) and about three times that among low-mass
stars (Preibisch et al. 1999).   

Binaries were taken into account in our synthetic HR diagrams, considering the
mass ratio distribution and a  percentage 
of double stars varying with mass, as described in section 3.2.

\subsection {Selection of the Hipparcos sample}

Our sample of stars  for the local solar neighborhood taken from the Hipparcos 
Catalogue
has been selected according to the following requirements:
  
\noindent
(i) All main sequence or evolved stars within 50 pc of the Sun with 
a relative parallax accuracy better than 10 per cent are considered
(corresponding magnitude completeness limit is down to $M_V=4.5$).
The standard error for (B-V) is required to be less than 0.025 mag, so that
the observed data are actually reliable for the comparison with theoretical
models. 

\noindent
(ii) From this sample we removed stars belonging to the Hyades cluster 
according 
to Perryman's et al. (1998) list, as the cluster may represent a local 
peculiarity with respect to the average solar vicinity population. 

\noindent
 The sample thus derived with these constraints contains 1844 stars and we 
consider 
the corresponding HR diagram sufficiently populated in the regions of figure 1a
for our analysis of the star formation history.

\noindent
Double and multiple systems were taken into account according to the 
information given in the catalogue (ESA SP-1200, 1997).  
There are resolved systems with or without separate entries for the components,
but most of the systems are unresolved. Dealing with all the many cases of
multiplicity is very complicated, thus we will take into account binarity with
a few simplifications, being aware that there are also some selection 
effects (like undetected and/or  undetectable binaries for observational
limits) which we are not able to take into account. As a starting point 
we identified the number of binary systems  in each of the different HR diagram
 regions 
defined in figure 1b. Then we used this percentage of binaries  when 
computing the synthetic HR diagrams.

\section{Comparing the observations with theoretical models}

\subsection {Preliminary technique and results}

Comparing the colour-magnitude diagram (CMD) of a stellar population  
with theoretical isochrones or synthetic HR diagrams is a technique
to derive information on the age , chemical composition, SFR and IMF of the 
considered population.
By simply superimposing a few selected isochrones with  solar chemical 
composition
on the observed HR diagram of the Hipparcos sample, we can derive some   
preliminary insights for the solar vicinity: 

\noindent
$\bullet$  star formation began about 10 Gyr ago: $T_i=10$ Gyr. This is 
determined from the red envelope
of  subgiant and giant branch stars.
Jimenez et al. (1998)   suggested a minimum value
of the order of 8 Gyr  for the galactic disk, derived essentially by isochrones 
fit to the subgiant region of the Hipparcos sample. 
Combining kinematic information to local color magnitude diagram, Binney et al.
(2000) deduced a solar neighbourhood age of $11.2 \pm 0.75$ Gyr.
Our evaluation of 10 Gyr is in between these two age determinations and 
is supported also by the results of the age determination of open clusters by
Carraro et al. (1999), who suggest an age of about 9-10 Gyr for the Galactic
disc.
Due to  uncertainties in stellar
evolutionary models  we cannot place an  error in our age estimate. 
In section 7.3 we  report 
on the effects of changing the initial age $T_i$ ;

\noindent 
$\bullet$ the star formation stopped around 100 million years ago: $T_f=0.1$ Gyr;

\noindent 
$\bullet$ the solar neighborhood cannot be described by a single chemical
composition.
It is evident from the width of the main sequence band and also from the 
color extension of the horizontal branch clump that there is an extended 
range of chemical composition of the observed stars. From the comparison 
between isochrones with different metal content (Bertelli et al. 1994) and
the observations we evaluate that the majority of the stars must have a chemical
composition in the range 0.008$\leq$Z$\leq$0.03. This choice of the chemical 
composition
is supported by the results of extensive spectroscopic observations of 
selected nearby F and G stars by Edvardsson et al. (1993).

\subsection{The synthetic HRD technique }

The analysis of the CMD of our selected sample  relies on 
the synthetic HR diagram technique  (ZVAR; Bertelli et al. 1992, Vallenari 
et al. 1996ab, 
Aparicio et al. 1997ab, Gallart et al. 1996, 1999).
It requires the following information:

\noindent
$\bullet$ The shape of the star formation rate.
We will use  two very simple hypotheses  shown
in figures 2a and 2b  where the vertical scale is arbitrary. 
The absolute values of the SFR scale will be fixed from  the
total amount of mass  converted in stars (in solar masses and computed for 
each synthetic model). The normalization imposed to the models requires the same
 number of MS stars as in the Hipparcos sample  (1658 stars more luminous than 
$M_V = 4.5$).  
In section 8  we derive the local mass density.

In the following we keep the initial time  $T_i$ (at which star formation
began) and the final time $T_f$ (at which star formation ceased) 
fixed at the values suggested by the preliminary results (section 3.1).

The two hypotheses for the star formation rate are:

a)  The {\em const-const} model, which is 
a  combination of two steady periods with a discontinuity at time
$T_b$. $T_b$ is allowed to vary  between $T_i$ and $T_f$. The ratio 
between the SFR at the age $T_f$ and that at $T_b$ is represented by the 
parameter $I_b$ and can also vary (fig. 2a). 
The model where the SFR rate is constant
over the entire time interval is a particular case  ($I_b$=1).

b) The {\em var-var} model represents the case in which the SFR 
increases by a factor of three  during the first time interval from
$T_i$ to $T_b$ and then  the change of the  SFR  slope  is 
characterized by the choice of the parameter $I_b$
( fig. 2b).

We considered  possible solutions varying the parameters $T_b$ and  $I_b$, 
in our analysis (section 5 and 6).
We also checked  the influence for different values of $T_i$ (section 7.3).

It is evident that the {\em var-var} model can describe many different SFR 
shapes from  the Milky
Way chemical evolution model by Chiappini et al. (1997), where the SFR
first increases and then decreases to the present age, to cases where the 
rate is continuously increasing or decreasing.  
{\bf Our method can recognize  broad trends in the star formation rate
during this long time interval (10 Gyr ago), but it is not useful to
trace fluctuations of short duration}.

\noindent
$\bullet$ The initial mass function (IMF). 

The IMF is defined by
\begin{equation}
 dN \quad \propto \quad M^{-x}dM 
\end{equation}

\noindent
for which we initially assume the classical Salpeter law
with $x=2.35$.  We also considered other power law values (section 7.1).

\noindent
$\bullet$ The relative lifetime spent by stars in each elemental area of the 
HRD.
This is derived by interpolation on an extended set of masses and chemical
compositions from the Padova library of stellar models (see references of
the whole set of models in Bertelli et al. 1994).

\noindent
$\bullet$ Chemical composition.  Edvardsson 
et al.(1993) and Ng and Bertelli (1998) did not find an age-metallicity 
relation for field stars , but a large scatter in metallicity ($0.008 < Z <
0.03$).
Then in each simulation the metallicity
is stochastically varied from star to star within this range.

\noindent
$\bullet$ Binaries. In the simulations we adopted the percentage of binaries 
per magnitude interval as obtained from the Hipparcos sample,
according to the identification and/or flags in the catalog 
(decreasing from 0.70 for the more massive primaries to about 0.27 at the
faint limit at $M_V = 4.5$). 
As far as the mass ratio q of the system is involved, we took into account 
two different observational results ; first, according to Trimble (1990)
the distribution N(q) is described by a power law ($q^{-1}$) over the range
$q=0.1-1.0$; second, according to Mermilliod et al. (1992) there is a flat 
distribution between 0.4 and 1.
Of course this second distribution, giving no information about the behaviour
for $q<0.4$, suffers from the limit of neglecting the low mass-ratio 
companions, whose effective number nobody knows. On the other hand, if the 
primary is rather more massive than the companion, its magnitude and colour 
are almost the same as those of a corresponding single star.

\section{Criteria for best solutions }

In order to obtain the best solutions for our models
we separately compared the data for the main sequence stars (MS) and evolved 
stars (Red), which are identified by the separation  line shown in the color 
magnitude diagram of figure 1a.

\noindent
{\bf The MS  region.} 
The luminosity interval spanned by the main sequence stars is divided into
ten bins (zones 4-13 in fig 1b). For both the 
particular shapes of the SFR ({\em const-const}, {\em var-var}) many 
combinations of $T_b$, $I_b$ and $x$ have been examined.
For every model 20 simulations of the CMD have been generated using  the
observed number of MS stars as the normalization parameter. 
We assume that the SFR (for given IMF) of the model reproduces the features
of the observed sample with a reasonable agreement, when the two following 
criteria are satisfied:

\noindent
{\bf MSa)}  the value of
  
\begin{equation} 
   \chi_{MS}^{2}= \sum_{i=1}^n \big[(N_{obs}- \overline N_{model})^{2}/\overline N_{model}\big]_i 
\end{equation}

\noindent
must be minimized. $N_{obs}$ and
$\overline N_{model}$ are the number of stars in the i-th MS zone  of 
the HR diagram, from the observations and from the model (averaging over the 
20 simulations) respectively, and  n is the total number of MS zones (n=10). 

The probability distribution function for $\chi_{MS}^{2}$ 
 is tabulated (Bevington 1969). In our case there are  seven degrees of
freedom. If we require that the probability of the $\chi_{MS}^{2}$  is greater 
than
P=0.1,  we deduce that the acceptable minima must be 
below a critical value of 12.
For greater values of $\chi_{MS}^{2}$, the probability of 
obtaining such a large value of $\chi_{MS}^{2}$ with the correct SFR is 
smaller than 0.1,
indicating that the SFR actually used may not be appropriate.
We will test this method with  numerical simulations presented in the next 
section.

\noindent
{\bf MSb)} For the second MS criteria the Kolmogorov-Smirnov  test (K-S) is 
used which can  discard
models for which the agreement  between the observed and simulated CMD
must be excluded with high probability We have discarded models for which 
the probability P coming from this test is lower than P=0.1 (see Press et al.
1986 for more information on the K-S test). 
The 20 simulations were considered all together for the K-S test.

\noindent
{\bf The RED region. }
This region has been divided in three zones. Zone number 1
in figure 1b contains 
all red stars brighter  than $M_v=1.5$. Since  the majority of these
stars are in the central He-burning phase we  refer to them 
using the suffix $He$.
The other two "red" zones (zone 2 and zone 3 in figure 1b) have been selected   
in such a way that the ratio  between the number of stars in the 
zone 2 and 
that in the zone 3 ($R_{red}$) is indicative of the spread of the evolved
stars below the luminosity limit $M_V=1.5$.  The majority of the simulations 
shows that these stars are concentrated in a narrow band inside the zone 3,
and no stars in zone 2 ($R_{red}$=0).   
A further significant indicator of the star distribution is the ratio 
$R_{He/MS}$ between  the number of stars in the He-burning phase, $N_{He}$, 
and that in the main sequence phase, $N_{MS}$. {\em This ratio is not very 
sensitive to the details
of the SFR, while it depends on the IMF parameter $x$ and on the stellar 
structure
(for example on the overshoot parameter $\lambda$).} In order to identify the  
models which give a satisfactory representation of the red part of the HR 
diagram we consider the following criteria:

\noindent 
{\bf Ra)} the quantity $\chi_{red}^{2}$ (the same meaning of $\chi_{MS}^{2}$, 
but relative to the red zones 1, 2 and 3) must be minimized. In this case, 
because of the small  number of zones, the degrees of
freedom are zero and the statistics of the distribution of 
$\chi_{red}^{2}$ is not 
known; we have done numerical experiments suggesting that the minimum must be 
below the value $\chi_{red}^{2}$=5;

\noindent
{\bf Rb)} the observed $R_{red}$ must be within $\pm 1\sigma$ of the mean
value obtained by the 20 simulations, where $\sigma$ is the standard deviation;

\noindent
{\bf Rc)} the observed ratio  $R_{He/MS}$ must be within $\pm 1\sigma$ of the
mean value obtained by computations.

\section{Test of the method}

In this section we test the reliability  of the criteria  adopted 
in evaluating the results. We proceed in a 
similar way as in Gallart et al. (1999).
We compute a stellar population with a given  star formation rate and 
known input parameters (i.e. $T_i$=10 Gyr, $T_f$=0.1 Gyr, metal content Z 
chosen 
stocastically in the range 0.008-0.03, IMF parameter  $x$=2.35, binarity
percentage as from Hipparcos data and distribution of mass ratio as 
in Trimble (1992)). We used  the same number of stars  as  the Hipparcos sample
since we considered this population as the "observed sample".
The particular SFR adopted for this test is {\em constant} over the entire 
time interval. 
 
For a fixed value of $I_b$ (6 cases are investigated, namely $I_b = 0.2, 0.6, 
1.0, 1.5, 2.0, 2.5$), the corresponding values of $\chi_{MS}^2$ and 
$\chi_{red}^2$ are computed for $T_b$ varying from $T_i$ to $T_f$, spaced by
1 Gyr. We expect to single out  the values of $T_b$ and $I_b$ 
corresponding to a minimum  of the $\chi^2$ functions,
and requiring that the previously defined criteria for the Main Sequence and
the Red region  be satisfied.

We begin with the {\em const-const} case (see fig. 2a) for the SFR.
In fig. 3a the results relative to the 
function $\chi_{MS}^2$ are presented. The continuous lines with
increasing thickness correspond to increasing values of $I_b$ (0.2, 0.6,
1.0). The dotted, dot short-dashed and dot long-dashed lines refer
respectively to the values $I_b$=1.5, 2.0, 2.5.
The line corresponding to $I_b$=1.0, which means a constant SFR
(independent of $T_b$), fluctuates between 6 and 10, below the critical value 
$\chi_{MScrit}^2$=12 
and around a horizontal mean line; 
this mean line represents a lower value with respect to all other cases
considered for $I_b$. 
For the specific case $I_b$=1 all models satisfy the K-S test.
In  the corresponding figure all models  which satisfy the K-S test  are 
indicated with a black square . 
The main result is that a constant SFR was recovered. 
However there is some  degeneracy.  
In fact, when $T_b$  approaches $T_i$, 
several other curves (the curves $I_b$=0.6 and 1.5) reach values of 
$\chi_{MS}^2$ 
in the range 6-10 and the method is unable  
to disentangle between these solutions.  In any case {\bf the 
result of the analysis with
a const-const model is consistent with a SFR constant during
the majority of the lifetime of the system} (at least during the last 
8 Gyr).

 The results from the red part of the
HR diagram  support our previous analysis. 
In fact, as shown in fig 3b,
there are  two curves which fluctuate around  a straight line  below the 
value 
$\chi_{red}^2$=5, the curve $I_b$=1 (constant SFR) together with the 
curve  $I_b$=0.6, so that the condition {\bf Ra}) is satisfied.
They satisfy also the condition {\bf Rb}), (models indicated with a solid
triangle) and the condition {\bf Rc}) (indicated with a big open triangle).
We point out  that for almost all the values of 
$I_b$ and $T_b$ the condition {\bf Rc}) relative to the ratio  $R_{He/MS}$  
is satisfied (it is within $\pm 1\sigma$ of the "observed" value). This means 
that this ratio is insensitive to the particular shape of the SFR. 
The analysis for the {\em var-var} model confirms   
 a constant SFR at least during the last 8 Gyr.

The fact that different solutions exist which give the same information, that 
is a constant SFR during the majority  of the system lifetime, except some
indetermination at the beginning, should clearly  indicate that 
{\bf the capability
of the method is not the determination of the detailed shape of the SFR, but 
simply the general trend over the total life of the system}.
 
We  repeated the same analysis for two values of the IMF slope  ($x=1.35$
and $x=3.35$ (remind that the simulated observed sample was derived with 
$x=2.35$). In both cases the analysis did not give acceptable solutions.

\section{Application to the Hipparcos sample}

We now present our results using 
the Hipparcos sample where we adopted  the input parameters
already discussed at the beginning of section 5. 

\smallskip
\noindent
{\bf  a) Const-const Models}
 
In fig 4a
we present, as a function of $T_b$ and $I_b$,   the 
trend of $\chi_{MS}^{2}$ for the main sequence stars.
The black squares  in the figure mark those models
whose probability P  from the K-S test is higher than 0.1. 
  From the figure we notice   
that the three curves characterized by $I_b>1$ ($I_b$=1.5, 2.0, 2.5) reach
 minimum  values of $\chi_{MS}^2$  below the critical value
12. The corresponding solutions are  respectively 2.5 Gyr for $I_b$=1.5, 
4.5 Gyr for 
$I_b$=2.0 and finally 5.5 Gyr for $I_b$=2.5. For the models around the minima 
also the K-S test gives acceptable results.

In fig 4b the curves $\chi_{red}^2$ (relative to the red part of the HR diagram)
corresponding to the same models of fig 4a are drawn. 
There is a broad minimum for $I_b$ = 2.0 and 2.5, with  low values of
$\chi_{red}^2$ with respect to all  other curves, even if this minimum
is of the order of 25 (5 times the expected value). Consequently the condition 
{\bf Ra}) is not satisfied; the age range is 
slightly shifted to younger 
values with respect to those obtained for the main sequence.    
Furthermore no models, in correspondence of the minima, satisfy conditions
 {\bf Rb}) and {\bf Rc}).

\smallskip
\noindent
{\bf c) Var-var Models}

In fig 5a $\chi_{MS}^2$ versus $T_b$ is shown.
There are several models which could fit the observed distribution of the MS
stars. 
 In the range 6.5-9.5 Gyr the curves $I_b>1$ attain a minimum or fluctuate
around low values of $\chi_{MS}^2$ (below the critical value 12 as shown in
the figure)  where  the requested condition on the K-S test is also satisfied.
In the range 1.5-3.5 Gyr, solutions 
are possible also for the case $I_b$=1; this means that the SFR was increasing 
by a 
factor of 3 from the beginning until  1.5-3.5 Gyr ago and then remains constant 
to the present.
It could seem that the result of the analysis  is not very coherent, but 
all the models suggest an increasing SFR on average from the 
beginning up to now. For the red region we meet the
same difficulties already expressed in the previous case. In particular  
all the models which could be considered good solutions on the base of the Main
 Sequence
analysis, give  $\chi_{red}^2$ values (fig 5b) which are 4-5 times higher 
than the critical 
value obtained in the test of section 5 and in consequence, fail to 
comply with  the 
significant condition {\bf Rc}). In fact
the ratio $R_{He/MS}$ fluctuates around a value which is about
1.5 times the observed one for all the models.

With the IMF parameter x=2.35 and the particular shapes of the 
SFR adopted, we could not find solutions satisfying  the
required conditions for the MS and the red region at the same time.  
 
It is well known that there are significant uncertainties in stellar models
due to the treatment of convective mixing and the consequent derivation
of convective zones border (overshoot problem). These uncertainties 
affect the H and He-burning lifetimes and their ratio, as well as the
expected number of stars in blue or red regions of the HR diagram.
Our models (synthetic HR diagrams) are normalized to the observed number of
main sequence stars, so that if the adopted overshoot treatment is not
fully correct, the major consequences will appear in the red part of 
the HR diagram.
We  decide to give up the  constraints 
{\bf Rb}) and {\bf Rc}) and to keep the condition that $\chi_{red}^2$ 
must show the lowest values with respect to all considered models,
assuming that the Salpeter IMF is a reasonable choice and that the 
overshoot parameter $\lambda$ used in our stellar
models is not well calibrated,
With  these relaxed conditions for the red region ($\chi_{red}^2$ minimum of 
the order of 20-30) the best solutions for the SFR parameters $I_b$ and $T_b$   
are presented in Table 1. 

\section{Effects of changing the input parameters}

\noindent
\subsection {The  IMF parameter x}

\smallskip
We explore other values for the IMF parameter to look
for its influence on the results.  The cases $x$=1.35 and $x$=3.35 have been 
considered.  

\smallskip
\noindent
a) {\bf x=1.35}

\smallskip

The {\em const-const} model  presents a reliable solution for the MS region,
given by a deep minimum at $T_b$=1.5 Gyr, with the parameter $I_b$ equal 
to 0.6.
The red counterpart of the 
MS solution is completely unsatisfactory as $\chi_{red}^2$ is 
greater than 140 and the ratio $R_{He/MS}$ is about 2 times greater
than the observed value (its value fluctuates around 0.093 for all the
models and combinations of $T_b$ and $I_B$. We show figures  6a and 
6b  for the 
{\em const-const} model, which can be compared with x=2.35 in figure 4a.

\smallskip
\noindent
b) {\bf x=3.35}
\smallskip

Figures 7a and 7b  show the curves
$\chi_{MS}^2$ and $\chi_{red}^2$ as a function of $T_b$ for the {\em var-var}
model. 
$\chi_{MS}^2$ reaches a minimum higher than the critical value for $I_b$=2.5 
at  $T_b=1.5$ Gyr.
The corresponding $\chi_{red}^2$ reaches a 
low value (of the order of 5) representing an acceptable solution.
The most interesting result  is that the ratio 
$R_{He/MS}$  
remains around 0.05, very near to the observed value 
(0.047) for all the possible choices of the star formation rate,
as indicated by the big open triangles in figure 7b.

Even if for this model we have the best solution for the red region (in fact
all the conditions {\bf Ra}), {\bf Rb}), and {\bf Rc}) are satisfied), 
in accordance with the discussion in section 6 we decide to give more credit 
to the results from the MS region, that are at the limit of acceptance
 for this value of the IMF slope.

\noindent
We can conclude that in our opinion
the best founded results are those obtained for the x=2.35 case. 

\noindent
{\bf Degeneracy between IMF and SFR}

The degeneracy between the SFR and the IMF is evident from the previous 
discussion relative to the cases $x=1.35$ and $x=3.35$. As far as 
the MS is concerned, we find that a flat IMF ($x=1.35$) favours a decreasing
SFR as indicated by the solution with $I_b$=0.6 for the {\em const-const} model,
while a steep IMF ($x=3.35$) requires an increasing SFR characterized by
$I_b=2.5$ in the  {\em var-var} model.
The ratio $R_{He/MS}$ changes approximately of a factor two going from 
$x=1.35$ to $x=3.35$.
As soon as the uncertainties on
stellar evolutionary lifetimes are removed, the information from the HR 
diagram red region could partly  eliminate this degeneracy.   

\noindent

\subsection { Binary percentages and  mass ratio distributions} 

We analyzed the effects of changing the percentage 
of binaries on the models. We considered two cases: the percentage
of binaries is respectively 75\% or 50\% of the total number of stars 
independently of the visual magnitude.
The results are not very much  influenced by these changes. Also the effect 
of the
shape of the distribution of the mass ratio has been considered; in 
place of  Trimble's (1990) distribution (as used in all previous 
computations), a flat distribution between 0.4 and 1.0 (according to 
Mermilliod et al. 1992) has been considered and also in this case 
the effects on the results are unimportant. The  effects become considerable
only when the mass ratio of the binary sistem is  near to  unity, 
producing a flatter
SFR with respect to previous results. 

\subsection { Initial age $T_i$}

We considered  $T_i$=8 Gyr and $T_i$=12 Gyr. The adoption  of a
rejuvenated initial age ($T_i$=8 Gyr) has the effect of flattening the SFR. In 
fact we find a constant SFR as an acceptable solution, analyzing the sample 
with the 
{\em const-const} model. An older value of the initial age ($T_i$=12 Gyr) 
does not modify the general conclusions presented in section 6
for the {\em const-const} and {\em var-var} models: a star
formation increasing (in a broad sense) from the beginning up to now.

\section {The local star formation rate and mass density}

The Hipparcos data allow the assessment of the amount of matter in the 
local Galactic disk (Holmberg and Flynn, 2000), firstly because of the 
information on the kinematics 
and the vertical density distribution of stars, secondly because of the
measurement of the local luminosity function.
From a volume-complete sample of A and F stars 
Holmberg and Flynn (2000) derived an estimate of 0.095 $M_\odot pc^{-3}$ 
in visible disc matter. Other determinations of the 
local mass density are by Cr\'ez\'e et al. (1998), who estimated the total 
amount of gravitating
 matter $0.076 \pm 0.015 M_\odot pc^{-3}$, and by Pham (1997)
who derived a local dynamical density $\rho_0 = 0.11 \pm 0.01 M_\odot pc^{-3}$.

We must point out that all our previous discussions involved only the more
luminous part of the HR diagram ($M_V \leq 4.5$) and 
all tests with different values of the IMF slope used only stars with
masses greater than about 0.9 $M_\odot$.
The absolute scale of the SFR is determined by the total mass converted into
stars, computed imposing the same number of MS stars  
as in the Hipparcos sample and extending the HR diagram simulations 
down to the low limit of the main sequence (0.1 $M_\odot$)
for the given IMF.  
 
Two possible choices for the power law IMF parameter  have 
been considered: a) x=2.35 (Salpeter) in eq. (1)  over all the mass range, and  
b) two different IMF slopes depending on the mass range: x=1.3 between 0.1 and 
$ 0.5 M_\odot$ (from Kroupa et al. 1993) and the Salpeter slope for masses
greater than  0.5 $M_\odot$.
This second hypothesis is indicated with TS IMF.
In figure 8 we show the SFR per unit volume and time computed for the 
solutions in Table 1, taking into account low mass stars with the TS IMF
hypothesis. 
In Table 2 
we present the volume density derived for the two considered IMF cases 
for the same Table 1 models. 
The stochastic fluctuations of the total mass per unit volume, due to the 
normalization to the small number of observed stars, determine the standard 
deviations for the volume density, which is, at maximum,  $6\%$ of the 
values reported in Table 2.

The most interesting points are:

\noindent
i) Given the low mass IMF, the derived volume densities corresponding 
to different SFR models are in a very narrow range, 
confirming  a  common characteristic in all solutions: {\bf the star formation 
rate increases, in a broad sense, from the beginning to the present time}. 
The sentence "broad sense" is justified by the results in figure 8, as the 
SFR at the beginning was in all cases smaller by a factor 2-5 than the 
present one.    

\noindent
ii) The differences between the volume density derived
with a Salpeter law ($\sim 0.10$ $M_\odot/pc^3$) and that with  TS IMF  
($\sim 0.07$ $M_\odot/pc^3$) are significant and clearly due to 
the different IMF slope in the mass range from 0.1  to 0.5 $M_\odot$.
Due to the standard deviations in the volume density there is an error of
the same magnitude ($6\%$) for the SFR in figure 8.
This uncertainty is negligible with respect to the relative 
difference in the SFR value (of the order of $40\%$) corresponding to the 
two different IMF hypotheses. 

\noindent
iii)
Estimates of the local mass density of the ISM give  0.04 
$M_\odot/pc^3$ (Jahreiss et al., 1998,       
Cr\'ez\'e et al., 1998, Holmberg and Flynn, 2000  and references therein), 
but errors by a factor of two or more
are not excluded.
Our lower value of the stellar mass density ($\sim 0.07$ 
$M_\odot/pc^3$) is obtained for the  TS IMF case.
If $\sim 0.1$ $M_\odot/pc^3$ is the observed value 
of the dynamical mass density and 0.01 $M_\odot/pc^3$ the local mass density 
of brown dwarfs (Jahreiss et al. 1998),  filling the 
gap between the dynamical mass determinations and the mass density of stars
(luminous and brown) requires a drastic reduction of the ISM mass 
density of the order of $50\%$. In the
case of the dynamical mass obtained by Cr\'ez\'e et al. (1998) the 
reduction imposed to the ISM would be even more significant.

\noindent
iv) The high values obtained for the volume density with the Salpeter 
hypothesis for the entire mass range favours the TS IMF choice for the
low mass range ($0.5-0.1 M_{\odot}$). Also the discussion in point iii)
supports the TS IMF results.  
    
 As the vertical velocity dispersion relative to the galactic plane depends on
the age of the considered population, each population of given age is
characterized by different scale heights which increase with age.
Following  Hernandez et al.(2000) based on the  
results by Kujiken and Gilmore (1989), it is possible to derive the fraction 
of the total number of stars of age t over the disk thickness, 
relative to those inside a sphere of  given radius.
This procedure takes into account the details of the vertical disk force law
and the variations of velocity dispersion with age, which require a larger 
correction factor with age.

Knowing that fraction as a function of time from the models in figure 8 it is 
possible to derive the star formation rate expressed in $M_\odot/Gyr/pc^2$.
The results are shown in figure 9.  
The present SFR in the solar neighbourhood is of the order of 2 for all the
 TS IMF models. The observational estimates are between 2 and 10  
$M_\odot/Gyr/pc^2$, see G\"usten and Mezger (1982).
By integrating over time we obtain the total surface mass density in  
$M_\odot/pc^2$. The results are presented in Table 3 for the two considered 
IMFs. For the TS IMF case the total density of the stellar component is in 
the range 44-45 $M_\odot/pc^2$, in agreement
with recent evaluations ($50 M_\odot/pc^2$, Kuijken and Gilmore,
1989a; $48 M_\odot/pc^2$, Kuijken and Gilmore,1991; $49 M_\odot/pc^2$, Flynn
and Fuchs, 1994; $45 M_\odot/pc^2$, Englmaier and Gerhard, 1999; 
$48 M_\odot/pc^2$, Holmberg and Flynn, 2000).

\section{Discussion and conclusions}

We determined the star formation rate for the solar neighborhood
using the Hipparcos selected sample  and our theoretical stellar models.
Taking into account different SFRs and IMF slopes, we analyzed the main 
sequence region separately from the red region of the HR diagram.
The main results are:

1) All the accepted solutions, shown in Table 1, support a star formation
rate that was increasing, in a broad sense, from the beginning (10 Gyr ago)
up to the present time. 
 The range of stellar masses involved in the
Hipparcos sample (for which  $M_V  \ge 4.5$, corresponding to $M \ge
0.9-1.0 M_\odot$) requires the choice of the IMF only for the higher 
main sequence.
The IMF Salpeter slope  is adopted in Table 1 
solutions, since  flatter or steeper slopes are discarded by the comparison
with the observed sample.  

2) The volume mass
densities in Table 2 are strongly influenced  by the low mass main sequence IMF
and almost independent
of the different solutions for the SFR in Table 1.   
The volume density correspondent to the TS IMF case ($\sim 0.07 
M_\odot /pc^3$) requires a drastic reduction of the local mass density of the
interstellar matter in order to fill the gap between the dynamical mass
determinations and the mass density of the stellar components.
From the volume density of Table 2 we deduce the absolute
scale of the SFR. The solutions for the TS IMF case, shown in figure 8,
are in agreement with the present day value of the SFR 
($\sim 0.01 M_\odot /Gyr/pc^3$).
The hypothesis of a Salpeter IMF over the entire mass range seems definitely
ruled out.

3) The total surface mass density in $M_\odot /pc^2$, determined according
to Hernandez et al. (2000), is significantly different for the two IMF
hypotheses. For the TS IMF  case (two slopes for the IMF) the present  SFR 
and the total surface mass density of the stellar component
are respectively 2 $ M_\odot /Gyr/pc^2$ and about 45  $M_\odot /pc^2$,
values compatible with recent observational estimates.    

4) The ratio $R_{He/MS}$ could constrain the higher main sequence initial
mass function.
There is a degeneracy between the SFR and the IMF parameter x, and different 
combinations of them can satisfy the conditions requested for reliable 
MS solutions (section 7.1). As {\em the ratio $R_{He/MS}$  has the very 
important property of being almost 
independent of the
particular shape of the SFR and at the same time very sensitive to the 
parameter x of the IMF}, it would be conclusive for the IMF determination.
Unfortunately due to the still existing uncertainties on the input physics of 
the stellar
models (related to overshooting and/or mixing, for example) we cannot be 
confident on its theoretical value which depends on the ratio between the 
hydrogen and helium burning lifetime of the stellar models.   
In all the models of Table 1  the ratio $R_{He/MS}$ is a factor 1.5
greater than the observed one. If one excludes the possibility of some unforeseen selection effects,
 this means that the
evolutionary models are probably inadequate and present values of the
theoretical lifetime ratios are greater than the observed values derived from 
real star distributions in the HR diagram. 
As reviewed by Chiosi (1999) and Bertelli (2000),
stars whose turn-off mass is pertinent to the age range under consideration
(say from 0.1 to 8-9  Gyr) may be severely affected by the problem
of convective overshooting during both core H- and He-burning phases.
The subject is far from being settled. Furthermore,  it is also complicated by 
the fact that for stars in the mass range 1 to 1.5 - 1.6 $M_{\odot}$, 
during the MS phase, the convective core tends first to grow during a 
sizable fraction of the H-burning lifetime, and then to 
shrink. Whether or not  in these
circumstances convective overshooting from the H-burning core can grow to full
efficiency is still a matter of vivid debate (Aparicio et al. 1990, Rosvick 
et al. 1998, Chiosi 1999, Bertelli 2000 for a recent review). 
All this makes stellar models 
in this mass range still highly
uncertain, at least as far as detailed predictions are concerned. For these 
reasons we have given a different weight to the results coming from the MS 
analysis and to those coming from the red region.

\begin{acknowledgements}

We warmly thank Xavier Hernandez for helpful discussions and suggestions, 
Peter Stetson for a critical reading of the manuscript. 
Special thanks are due to Ron Canterna for his invaluable help to make more 
comprehensible the text. We are indebted to the referee for many useful 
comments that helped to improve the paper.

The Hipparcos sample has been selected making use of CDS facilities 
(Centre de Donn\`ees Astronomiques de Strasbourg, France).

This study has been financially supported by MURST.

\end{acknowledgements}

\clearpage

\figcaption[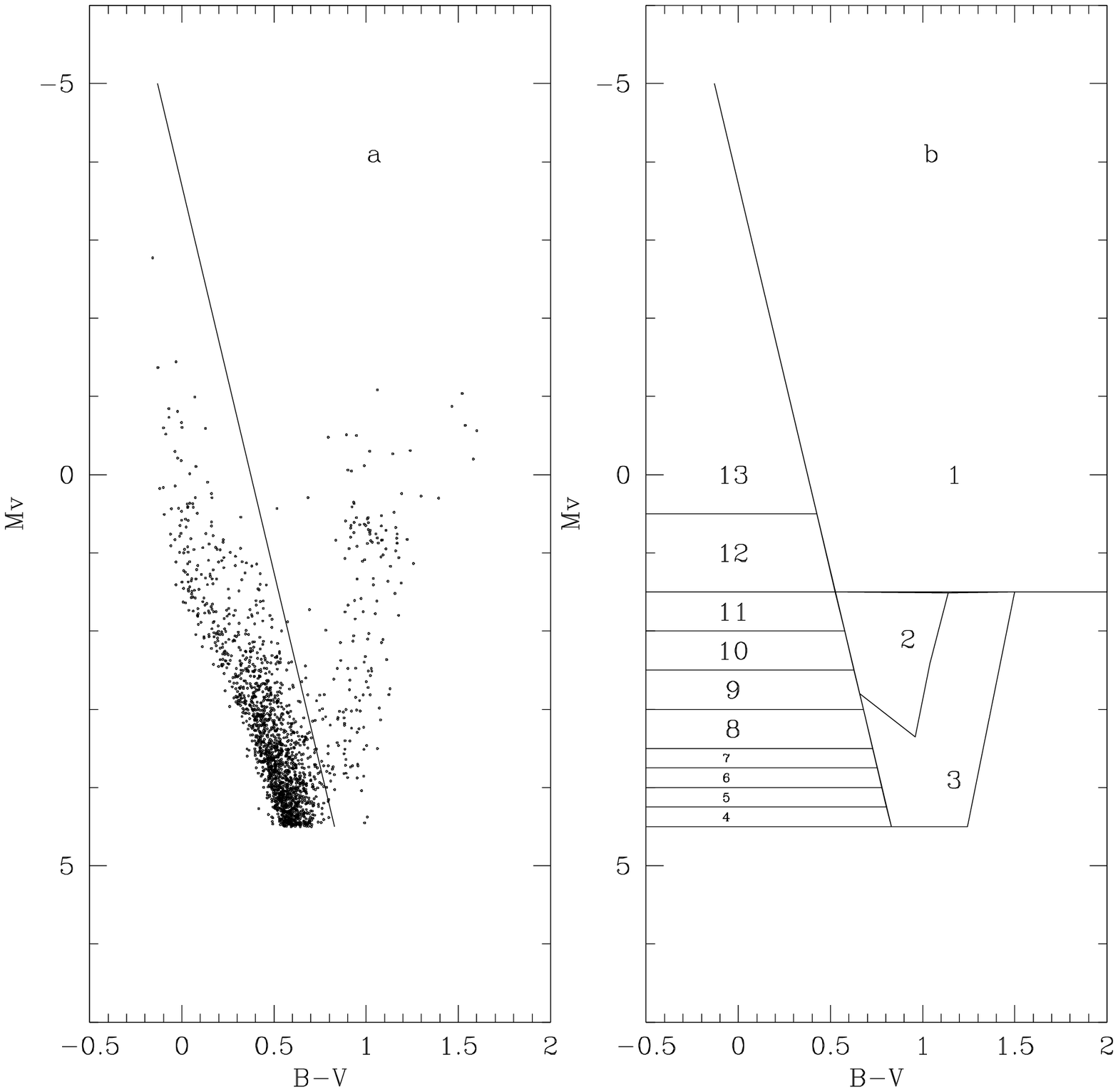]{ {\sl a}: the star sample 
selected from the Hypparcos catalogue with the criteria described in 
section 2.4. The transversal line separates MS from Red stars.
 {\sl b}: Star distributions: the HR diagram 
zones from 4 to 13 characterize the MS region, while
zones 1, 2 and 3 characterize the RED region. \label{fig1}}

\figcaption[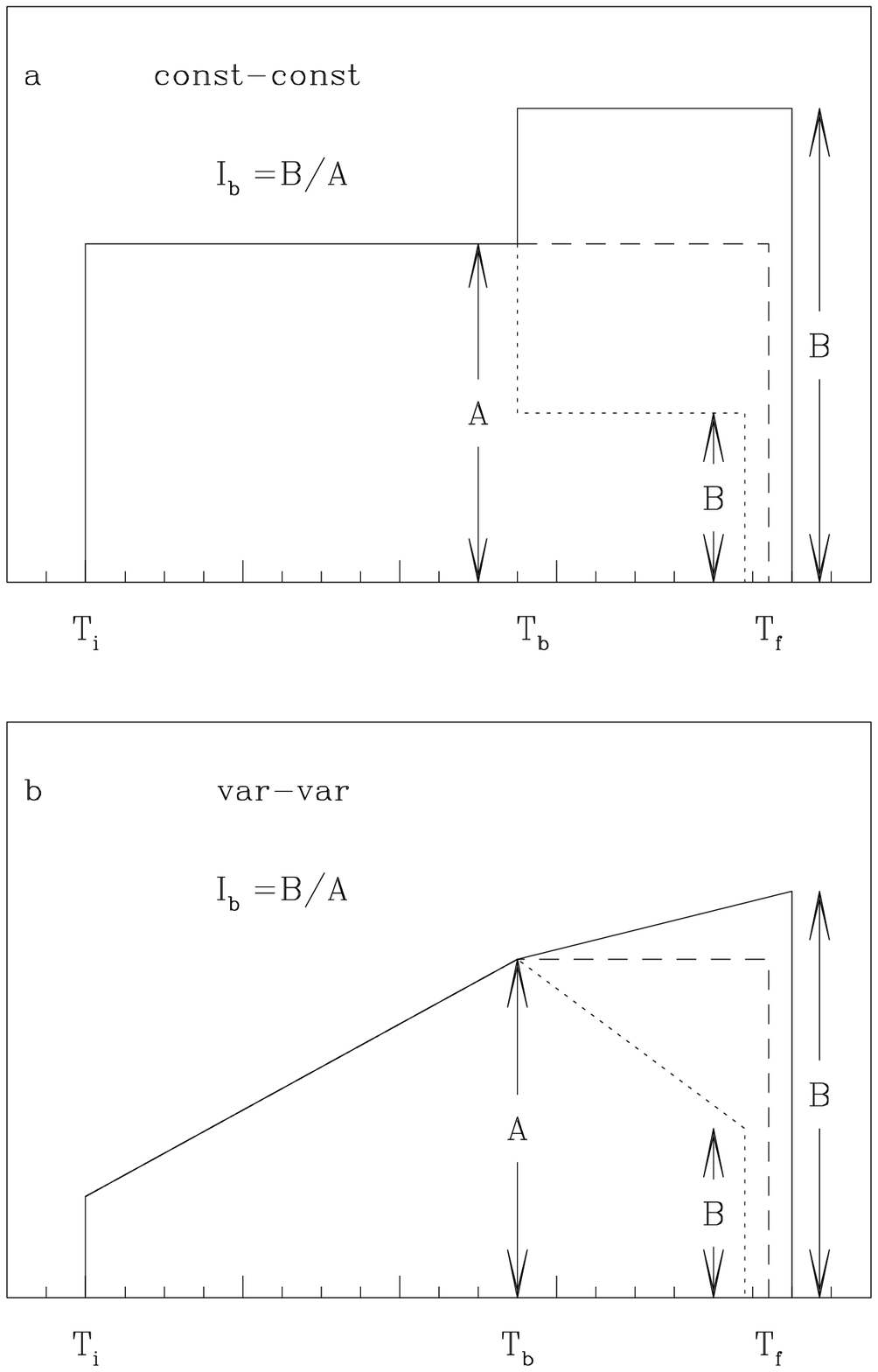]{ {\sl a}: Shape of the SFR, as described in 
section 3.2 for the {\em const-const}
model, where $T_b$ is the time of the rate change and $I_b$ is the ratio 
between the SF rate at the final time $T_f$ and the value before the change;
{\sl b}: the same as for panel a, but for the {\em var-var} model. In both 
panels the final time $T_f$ is the same for all the shapes, but is plotted
separately for the sake of clearness.
 \label{fig2}}

\figcaption[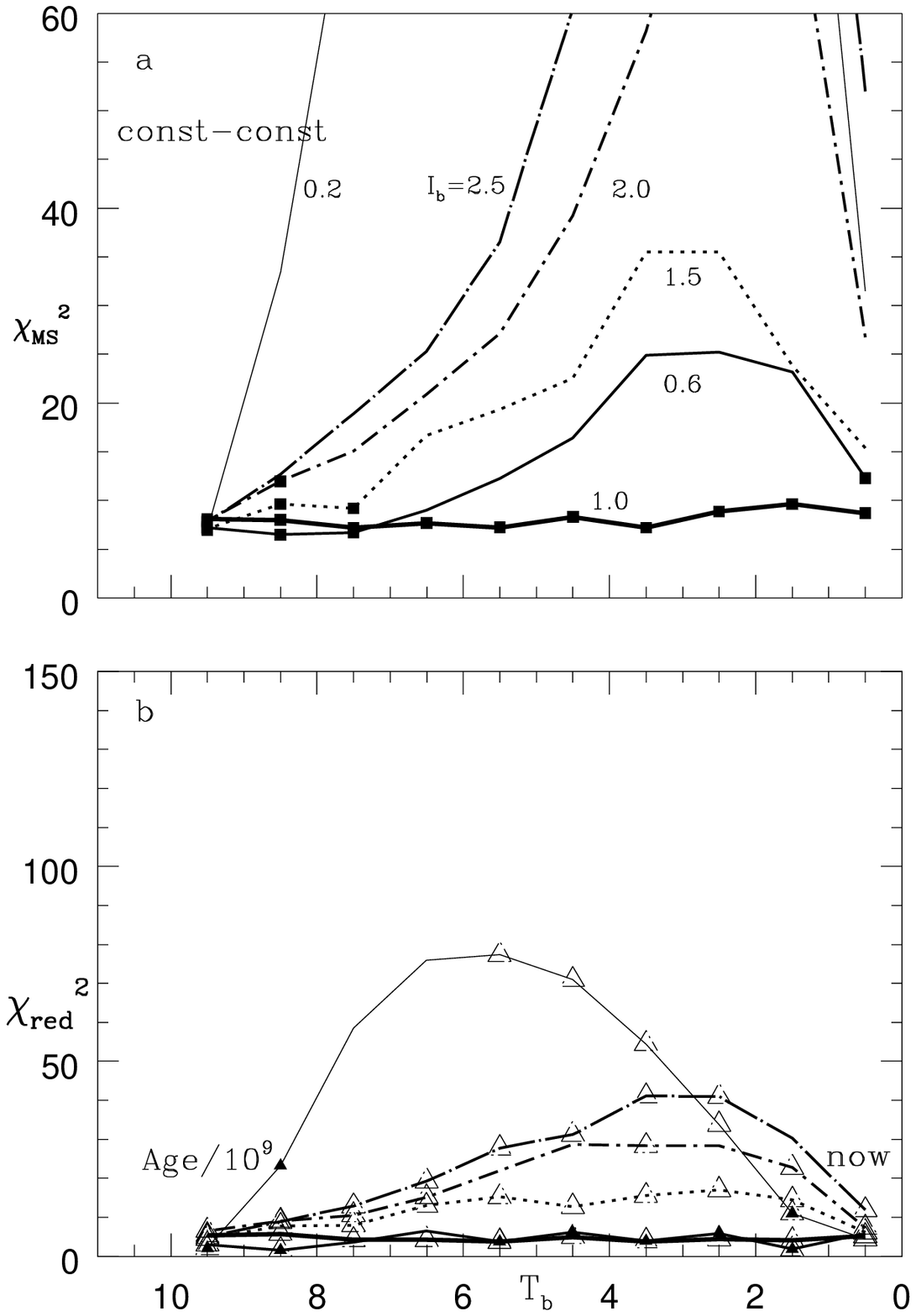]{ {\sl a}: $\chi_{MS}^2$ values computed for 
the SFR {\em const-const} model and with IMF x=2.35 to test the "simulated
sample" (produced with constant SFR and x=2.35). $T_b$ varies between $T_i$
and $T_f$, spaced  by 1 Gyr.  Different lines correspond to
different values of $I_b$, as shown. Solid squares indicate models whose
probability P from the K-S test is greater than 0.1
{\sl b}: the corresponding  $\chi_{red}^2$. Small solid triangles indicate 
that condition {\bf Rb}) is satisfied, big open triangles correspond to models
which satisfy condition {\bf Rc}).
 \label{fig3}}

\figcaption[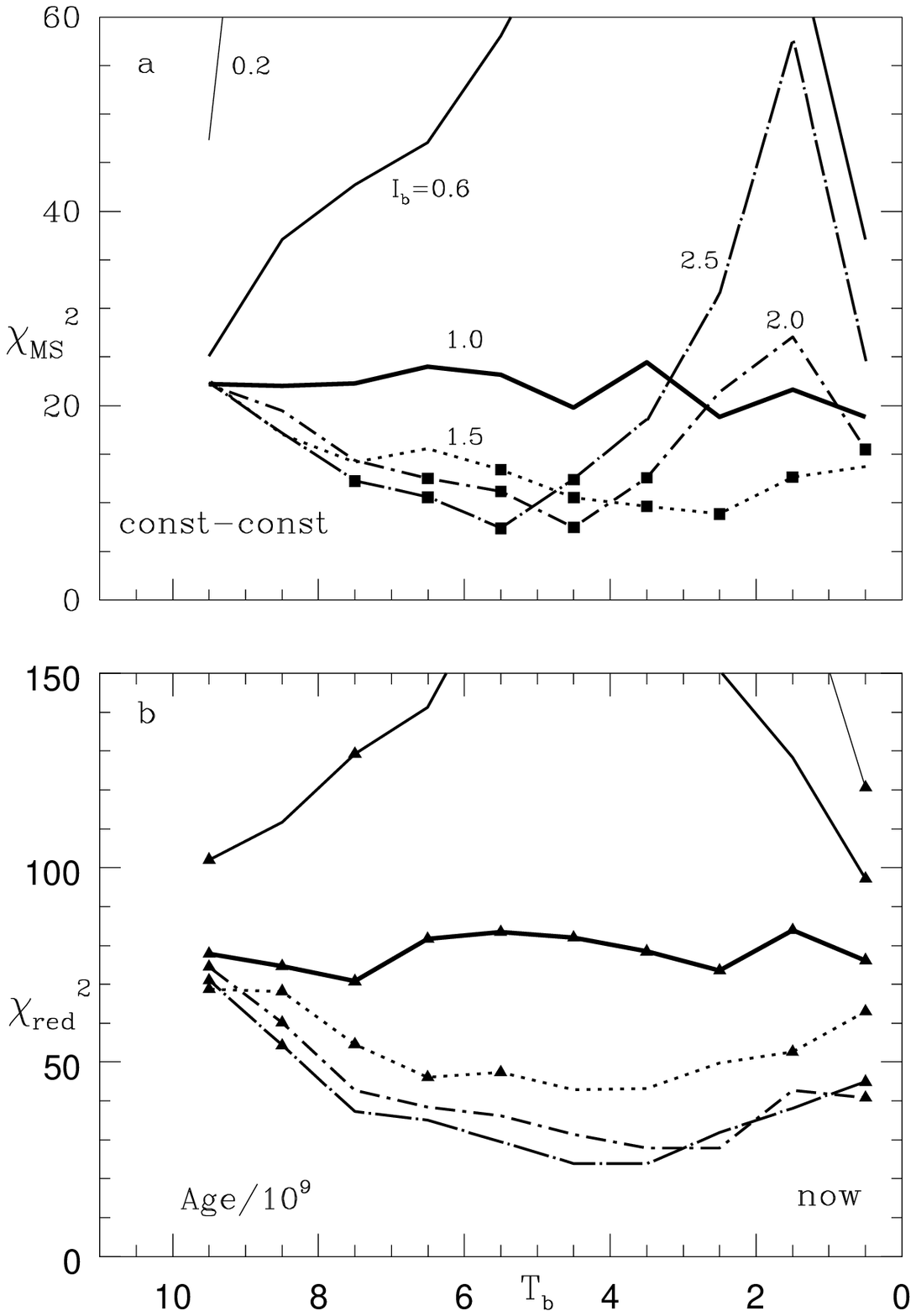]{ Analysis of the
Hipparcos sample with SFR {\em const-const} model and IMF x=2.35; {\sl a}:
$\chi_{MS}^2$ for main sequence stars distribution and {\sl b}: 
$\chi_{red}^2$ for red stars.
Symbols are the same as in figures 3a and 3b. \label{fig4}}

\figcaption[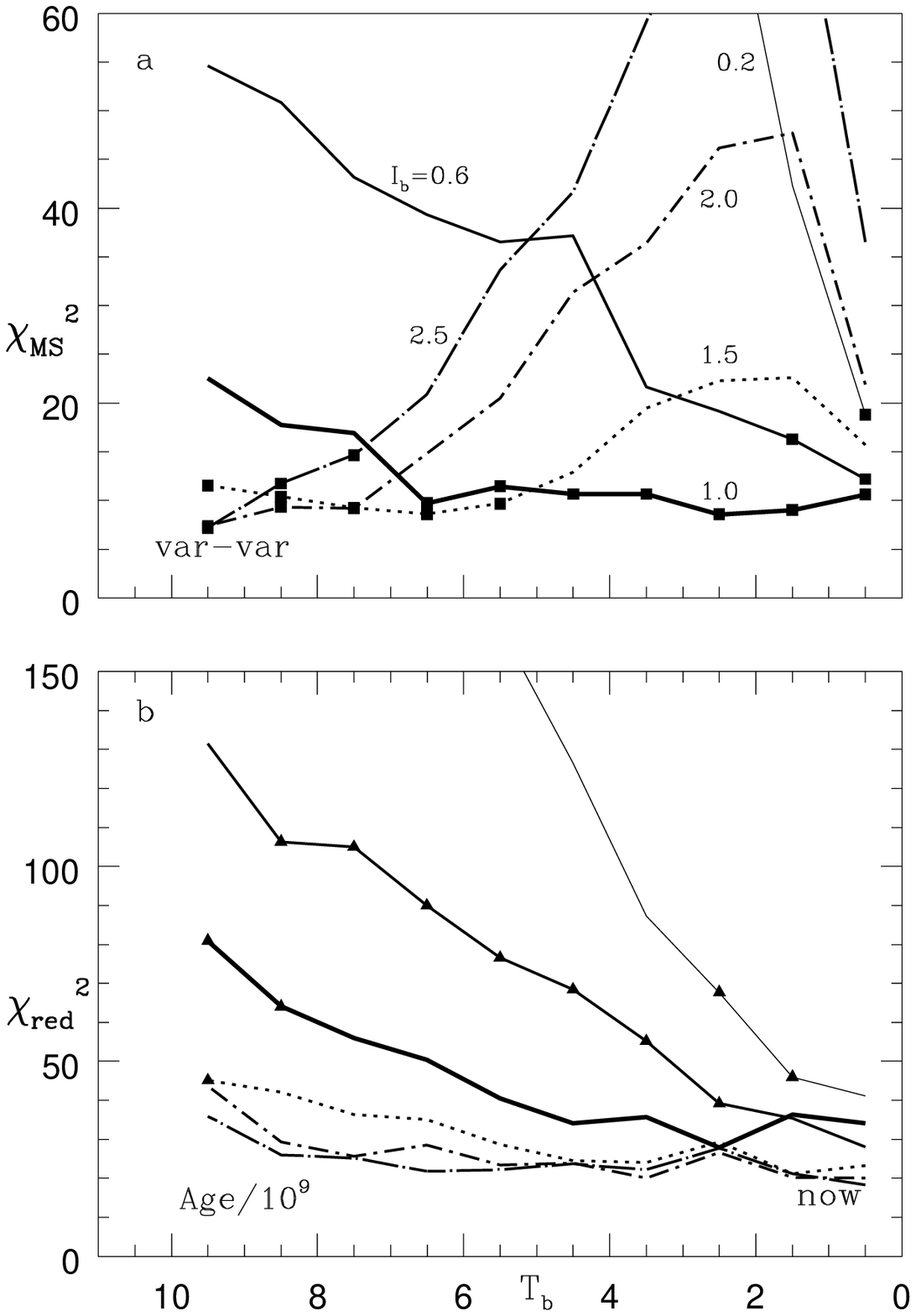]{ Analysis of the Hipparcos sample with
SFR {\em var-var} model and IMF x=2.35; Symbols are the same as in figures 
3a and 3b.
 \label{fig5}}

\figcaption[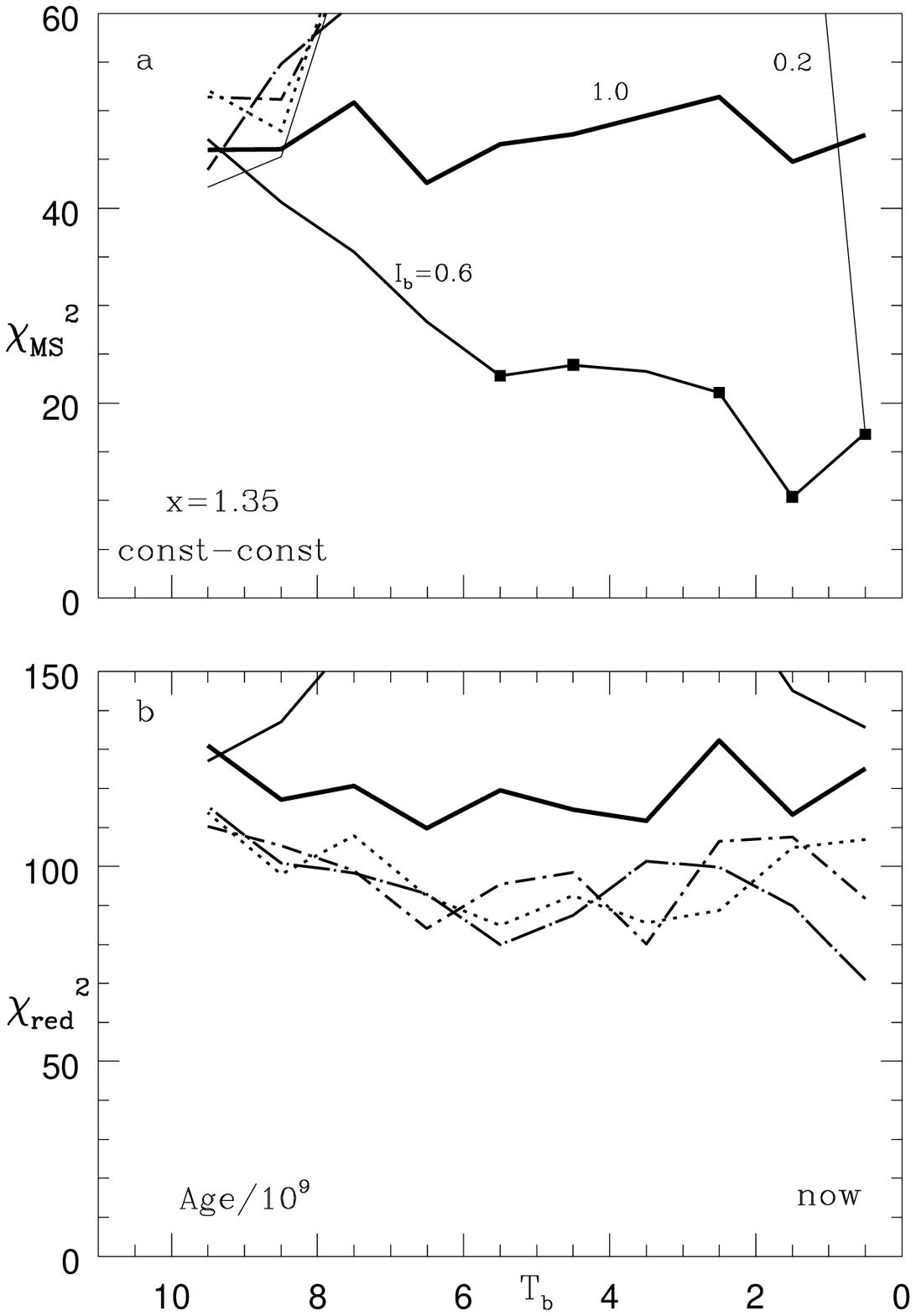]{ Analysis of the Hipparcos sample with SFR
{\em const-const} model and IMF {\bf x=1.35}. Symbols are the same as in 
figures 3a and 3b.
\label{fig6}}

\figcaption[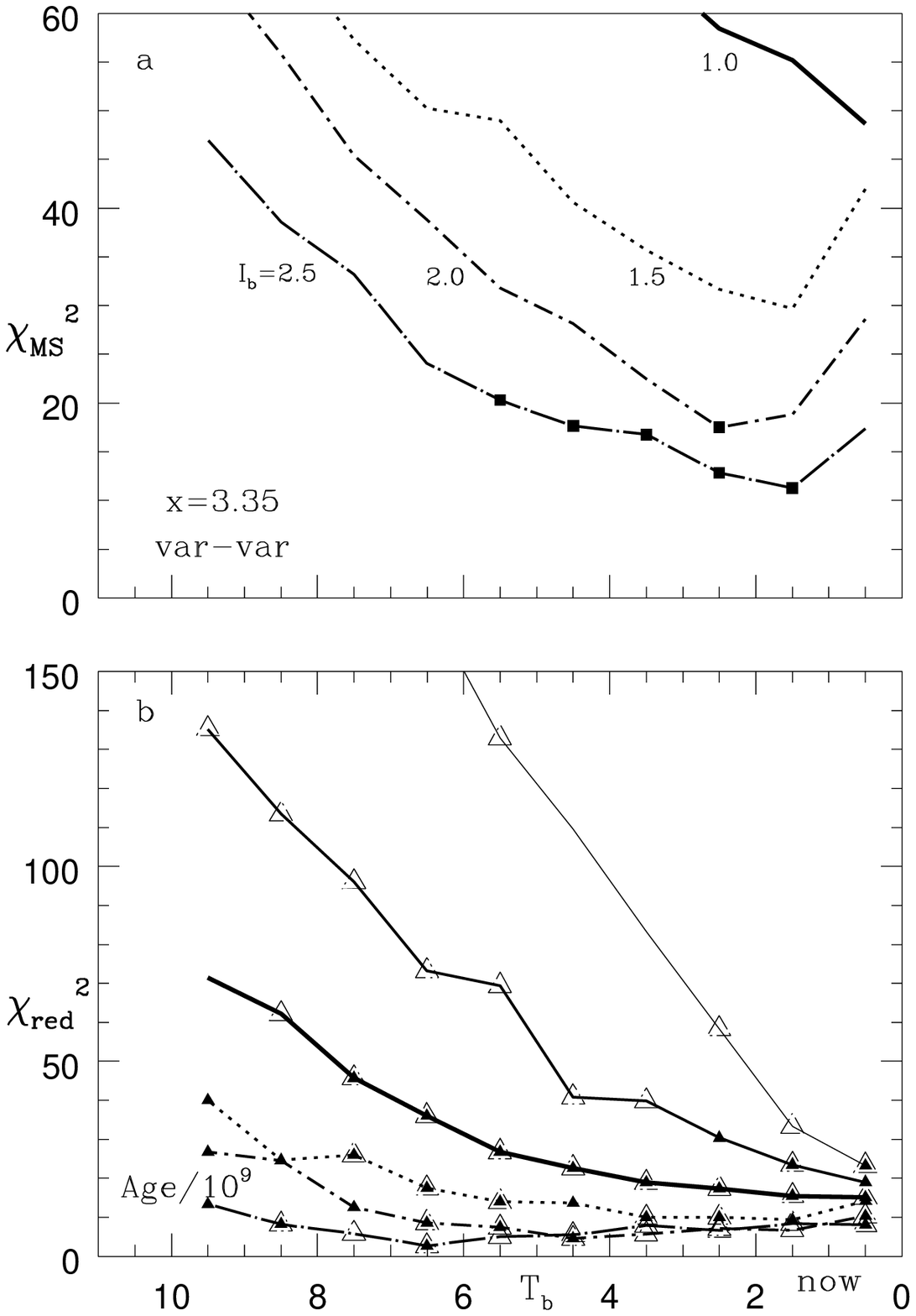]{ 
 Analysis of the Hipparcos sample with SFR
{\em var-var} model and IMF {\bf x=3.35}. Symbols are the same as in 
figures 3a and 3b.
\label{fig7}} 

\figcaption[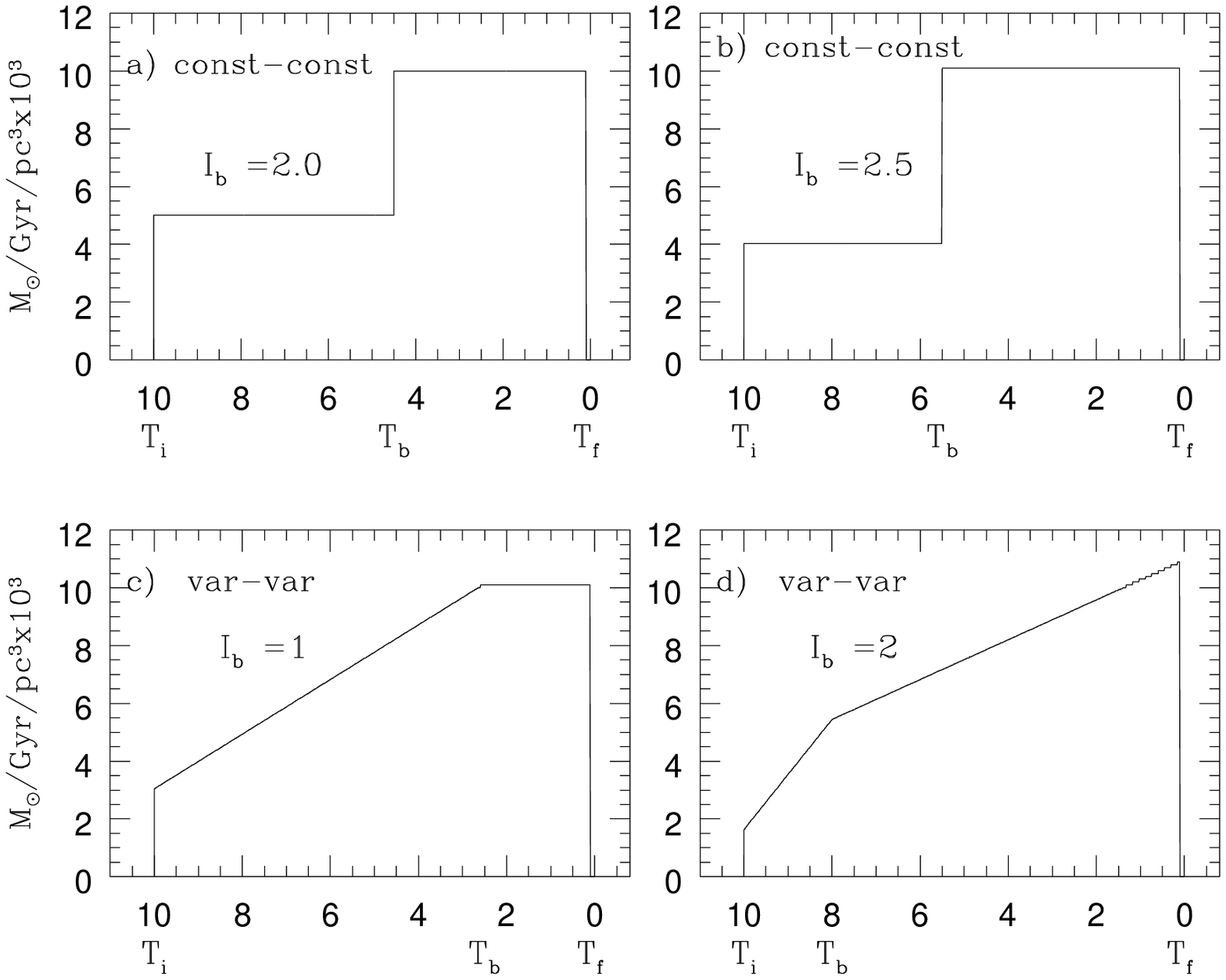]{ 
Star formation rate per unit volume in $M_\odot/Gyr/pc^3$ relative to the 
acceptable solutions
in Table 1. The SFRs were computed with initial mass function including 
the low mass stars  for the case TS IMF described in section 8.
Ages are expressed in Gyr.
\label{fig8}} 

\figcaption[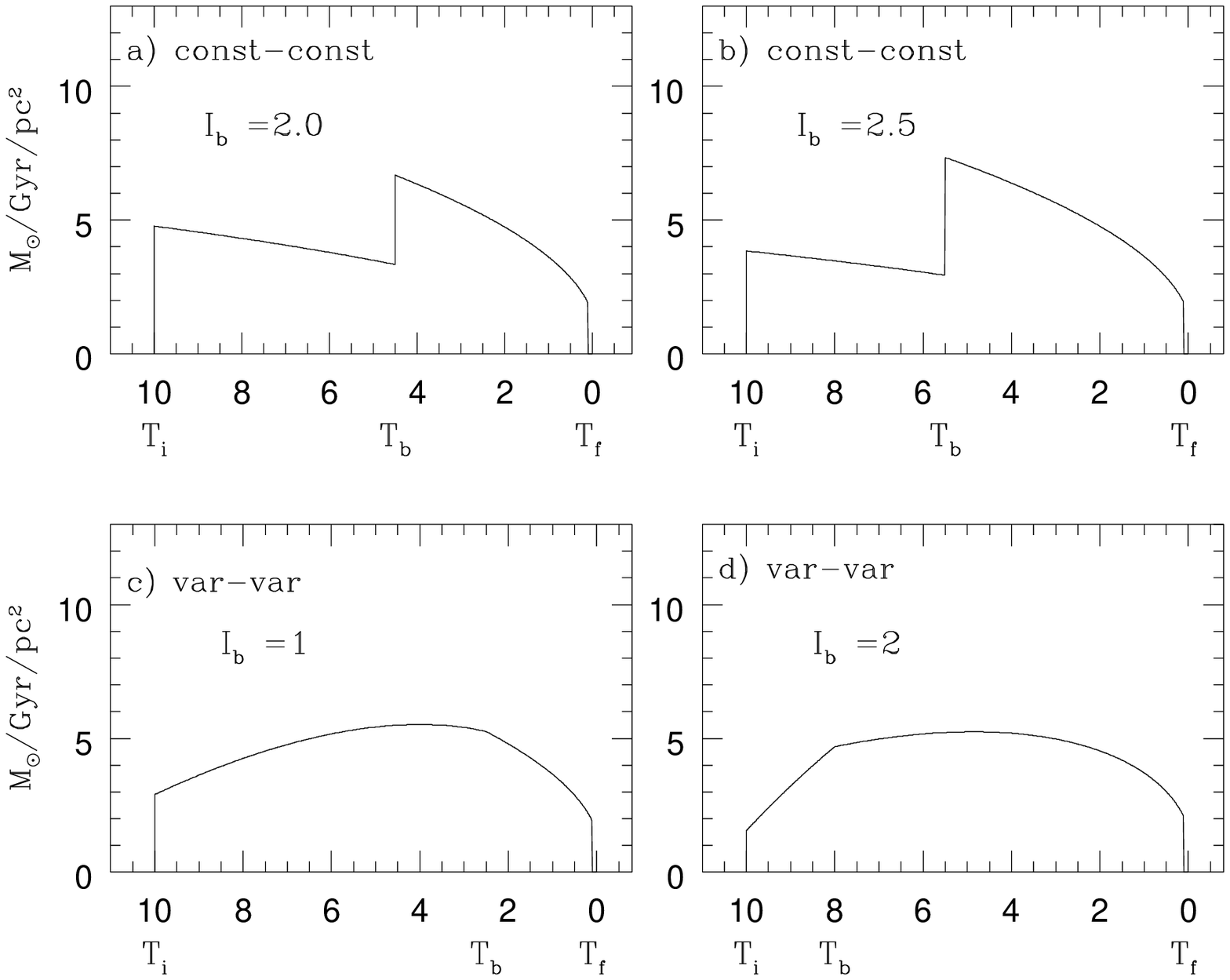]{ 
Star formation rates per unit surface area in $M_\odot/Gyr/pc^2$ derived from 
the models in figure 8,
accounting for  the fraction of stars of age t over the
disk thickness, relative to those inside a sphere of given radius.
Ages are expressed in Gyr.
\label{fig9}}


\begin{thebibliography}{}
\bibitem[Aparicio et al.(1990)]{} Aparicio, A., Bertelli, G., 
    Chiosi, C., and Garc\'{i}a-Pelayo, J. M.,  1990, \aap, 240, 262
\bibitem[Aparicio et al.(1997a)]{} Aparicio, A., Gallart, C., and Bertelli, 
    G., 1997a, \aj, 114, 669 
\bibitem[Aparicio et al.(1997b)]{} Aparicio, A., Gallart, C., 
     and Bertelli, G., 1997b, \aj, 114, 680 
\bibitem[Bertelli (2000)]{} Bertelli, G., 2000, in {\it Stellar Clusters and 
    Associations: Convection, Rotation and Dynamos}, eds.R. Pallavicini, G. 
    Micela and S. Sciortino, ASP Conf. Series 198, p 115
\bibitem[Bertelli et al.(1994)]{} Bertelli, G., Bressan, A., Chiosi, C., 
    Fagotto, F., and Nasi E., 1994, \aaps, 106, 275
\bibitem[Bertelli et al.(1999)]{} Bertelli, G., Bressan, A., Chiosi, C., 
    and Vallenari, A., 1999, Baltic Astronomy, 8, 271
\bibitem[Bertelli et al.(1992)]{} Bertelli, G., Mateo, M., Chiosi, C., 
    and Bressan A., 1992, \apj, 388, 400
\bibitem[Bertelli et al.(1997)]{} Bertelli, G., Nasi, E., Bressan, A., 
    and Chiosi, C., 1997, in {\it 
  Hipparcos: Venice 97}, eds. B. Battrick, M.A.C. Perryman, P.L. Bernacca,
    K.S. O'Flaherty, Noordwijk: ESA Publication Division, p. 501
\bibitem[Binney et al.(2000]{} Binney, J., Dehnen, W., and Bertelli, G., 2000,
    \mnras, subm. astro-ph/0003479
\bibitem[Carraro et al.(1999)]{} Carraro, G., Girardi, L., and Chiosi, C., 1999,
   \mnras, 309, 430
\bibitem[Chiappini et al.(1997)]{} Chiappini, C., Matteucci, F., 
   and Gratton, R.G., 1997, \apj, 477, 765
\bibitem[Chiosi (1999)]{} Chiosi, C.,  1999, in {\it Theory and Test of 
   Convective energy Transport}, 
   eds. A. Gimenez, E. Guinan, B. Montesinos,  PASP Conference Series, in
   press, p 1 
\bibitem[ Duquennoy and Mayor (1991)]{} Duquennoy, A., and Mayor, M., 1991,
    \aap, 248, 485
\bibitem[Edvardsson et al.(1993)]{} Edvardsson, B., Andersen, J., Gustaffson, B.,
    Lambert, D.L., Nissen, P.E., and Tomkin, J., 1993, \aap 275, 101
\bibitem[Englmaier  and Gerhard (1999)]{} Englmaier, P., and Gerhard, O., 1999,
    \mnras, 304, 512
\bibitem[Flynn and Fuchs (1994)]{} Flynn, C., and Fuchs, C., 1994,
    \mnras, 270, 471
\bibitem[Gallart et al.(1996)]{} Gallart, C., Aparicio, A., Bertelli, G., 
    and Chiosi, C., 1996, \aj, 112, 1950
\bibitem[Gallart et al.(1999)]{} Gallart, C., Freedman,W., Aparicio, A., 
   Bertelli, and G., Chiosi, C., 1999, \aj, 118, 2245 
\bibitem[Girardi (1999)]{} Girardi, L., 1999, \mnras, 308, 818
\bibitem[G\"usten and Mezger (1982)]{} G\"usten, R., and Mezger, P., 1982,
    Vistas in Astronomy, 26, 159
\bibitem[Jimenez et al.(1998)]{} Jimenez, R., Flynn, C., and Kotoneva E., 
    1998, \mnras, 299, 515
\bibitem[Jahreiss et al.(1998)]{} Jahreiss, H., Fuchs, B., and Flynn, C.,
    1998, \aap, 339, 40  
\bibitem[Hanson (1979)]{} Hanson, R.B., 1979, \mnras, 186, 875 
\bibitem[Hernandez et al.(2000)]{} Hernandez, X., Valls-Gabaud, D., and 
   Gilmore, G., 2000, \mnras, 316, 605 
\bibitem[Holmberg and Flynn (2000)]{} Holmberg, J., and Flynn, C., 2000,
    \mnras, 313, 209
\bibitem[Kroupa (1995)]{} Kroupa, P., 1995, \mnras, 277, 1507
\bibitem[Kuijken and Gilmore (1989)]{} Kuijken, K., and Gilmore, G., 1989a,
    \mnras, 239, 605 
\bibitem[Kuijken and Gilmore (1989)]{} Kuijken, K., and Gilmore, G., 1989b,
    \mnras, 239, 651 
\bibitem[Kuijken and Gilmore (1989)]{} Kuijken, K., and Gilmore, G., 1991,
    \apj, 367, L9 
\bibitem[Lutz and Kelker (1973)]{} Lutz, T.E., and Kelker, D.H., 1973, \pasp,
   85, 573
\bibitem[Mazeh et al.(1992)]{} Mazeh, T., Golberg, D., Duquennoy, A., and Mayor M.,
   1992, \apj, 401, 265  
\bibitem[Mermilliod et al.(1992)]{} Mermilliod, J.C., Rosvick, J.M., Duquennoy, A.,
   and Mayor, M., 1992, \aap, 265, 513
\bibitem[Ng et al.(1998)]{} Ng, Y.K., and Bertelli, G., 1998, \aap, 329, 943
\bibitem[Patience et al.(1998)]{} Patience, J., Ghez, A.M., Reid, I.N., 
    Weinberger, A.J., and Matthews, K., 1998, \aj, 115, 1972     
\bibitem[Perryman et al.(1995)]{} Perryman, M.A.C. et al., 1995, \aap, 304, 69
\bibitem[Perryman et al.(1998)]{} Perryman, M.A.C. et al. 1998, \aap, 331, 81
\bibitem[Pont et al.(1998)]{} Pont, F., Mayor, M., Turon, C., and VandenBerg, D.A.,
   1998, \aap, 329, 87
\bibitem[Press et al. 1986]{} Press, W.H., Teukolsky, S.A., Vetterling, W.T.
   and Flannery, B.P., 1986, in {\it Numerical Recipes in Fortran}, 
   Cambridge University Press, p 617
\bibitem[Preibisch et al.(1999)]{} Preibisch, T., Balega, Y., Hofmann, K-H.,
  Weigelt, G., and Zinnecker, H., 1999, New Astronomy, 4, 531 
\bibitem[Reid (1997)]{} Reid, I.N., 1997, \aj, 114, 161
\bibitem[Rocha-Pinto et al.(1996)]{} Rocha-Pinto, H.J., and  Maciel, W.J., 
   1996, \mnras, 279, 447
\bibitem[Rocha-Pinto et al.(1997)]{} Rocha-Pinto, H.J., and  Maciel, W.J., 
   1997, \mnras, 289, 882
\bibitem[Rocha-Pinto et al.(2000)]{} Rocha-Pinto, H.J., Scalo, J., Maciel, W.J.,
   and Flynn, C., 2000, \apj, 531, L115
\bibitem[Rosvick et al.(1998)]{} Rosvick, J. M., and VandenBerg, D.A., 
   1998, \aj, 115, 1516 
\bibitem[Schr\"oder (1998)]{} Schr\"oder K.P., 1998, \aap, 334, 901
\bibitem[Trimble (1990)]{} Trimble, V., 1990, \mnras, 242, 79
\bibitem[Turon et al. (1992)]{} Turon, C., et al., 1992, \aap, 258, 74  
\bibitem[Vallenari et al.(1996a)]{} Vallenari, A., Chiosi, C., Bertelli, G., 
    and Ortolani, S., 1996a, \aap, 309, 358 
\bibitem[Vallenari et al.(1996b)]{} Vallenari, A., Chiosi, C., Bertelli, G., 
   Aparicio, and A., Ortolani, S., 1996b, \aap, 309, 367 
\end{thebibliography}
\end{document}